\documentclass[12pt]{article}

\begin{document}

\begin{center}
{\Large\bf{}Modification of the Bel-Robinson type energy-momentum}
\end{center}

\begin{center}
Lau Loi So\footnote{email address: s0242010@gmail.com}\\
Department of Physics, National Central University, Chung-Li 320,
Taiwan
\end{center}

\begin{abstract}
For describing the non-negative gravitational energy-momentum in
terms of a pure Bel-Robinson type energy-momentum in a quasi-local
2-surface, both the Bel-Robinson tensor $B$ and tensor $V$ are
suitable. We have found that this Bel-Robinson type
energy-momentum can be modified such that it satisfies the Lorentz
covariant, future pointing and non-spacelike properties. We find
that these particular quasi-local energy-momentum properties can
be obtained from (i): $B$ or $V$ plus a tensor $S$ in a small
sphere limit, or (ii): directly evaluating the energy-momentum of
$B$ or $V$ in a  small ellipsoid region. (iii): calculate the
total energy using the Landau-Lifshitz pseudotensor in a small
ellipsoid, from Jupiter's tidal force to Io in Schwarzchild
spacetime, in an elliptic orbit.
\end{abstract}

\section{Introduction}
According to the Living Review article, Szabados said (see 4.2.2
in~\cite{Szabados}): ``Therefore, in vacuum in the leading $r^{5}$
order any coordinate and Lorentz-covariant quasi-local
energy-momentum expression which is non-spacelike and future
pointing must be proportional to the Bel-Robinson `momentum'
$B_{\mu\lambda\xi\kappa}t^{\lambda}t^{\xi}t^{\kappa}$.''  Note
that here $t^{\alpha}$ is the timelike unit vector and `momentum'
means 4-momentum. Previously, we believed that the Bel-Robinson
type energy-momentum was the natural choice and indeed the only
choice for describing the non-negative gravitational quasi-local
energy-momentum expression. However, we have now found that it is
not the case.

In the past, we thought there were only two gravitational
energy-momentum expressions that have the positive definite energy
(i.e., causal) since they give a positive multiple of the
Bel-Robinson type energy-momentum in a small sphere limit. They
are the Papapertrou
pseudotensor~\cite{1stSoCQG2009,Papapetrou,SoNesterCQG2009} and
tetrad-teleparallel energy-momentum gauge current
expression~\cite{Pereira,CJP}. We even had concluded that both the
Einstein and Landau-Lifshitz pseudotensors cannot guarantee
positive definite~\cite{1stSoCQG2009}, but now we discovered that
the Landau-Lifshitz pseudotensor ensure positivity while Einstein
does not. The motivation why we review the argument given by
Szabados~\cite{Szabados} is that we suspect there may exists a
relaxation such that the desired physical requirements can be
satisfied, i.e., the four-momentum are Lorentz covariant, future
pointing and non-spacelike. We find that the explanation given by
Szabados is necessary but not sufficient.

Positive gravitational energy is required for the stability of the
spacetime~\cite{Horowitz} and any quasi-local stress expression
which gives the Bel-Robinson type energy-momentum is the desirable
candidate. Moreover, evaluate the quasi-local energy-momentum
around a closed 2-surface, we can use the Bel-Robinson type
energy-momentum to test whether the expression can have a chance
to give the positivity at the large scale or not. Since negative
quasi-local energy guarantees negative for a large scale, while
positive quasi-local energy might have a chance for the large
scale. Checking the result for the gravitational energy in a small
regions is an economy way because the positivity energy prove is
not easy.

Basically, quasi-local methods are not fundamentally different
than pseudotensor methods~\cite{NesterPRL,Boot}. We will use the
pseudotensor to illustrate our modified quasi-local Bel-Robinson
type energy-momentum in three cases in section 3. Although
pseudotensor is an coordinates dependent object, it stills a
practical way to calculate the work done for an isolated system
from an external universe, e.g., tidal heating through
transferring the gravitational field from Jupiter to its satellite
Io~\cite{Purdue}. Tidal heating means when an external tidal field
$E_{ij}$ interacts with the evolving quadrupole moment $I_{ij}$ of
an isolated body, the tidal work per unit time is
$\frac{dW}{dt}=-\frac{c^{2}}{2}E_{ij}\frac{dI^{ij}}{dt}$, where
$I_{ij}\propto{}a^{5}_{0}E_{ij}$ and $a_{0}$ is the radius of Io.
This work rate formula is the same for the Newtonian energy and
general relativistic Landau-Lifshitz pseudotensor~\cite{Thorne}.
Tidal heating is a real physical observable irreversible process
that Jupiter distorts and heats up Io~\cite{Smith}, it should be
unambiguous of how one's choice to localize the energy, Purdue
used the Landau-Lifshitz pseudotensor to calculate the tidal
heating for Io in 1999~\cite{Purdue}. Two years later, Favata
examined different classical pseudotensors (i.e., Einstein,
Landau-Lifshitz, M$\o$ller and Bergmann conserved quantities) and
discovered the same tidal heating formula~\cite{Favata}. Moreover,
in 2000, Booth and Creighton modified the Brown and York
quasi-local energy formalism and obtained the same result for the
tidal dissipation formula~\cite{Boot}.

\section{Technical background}
The Bel-Robinson tensor $B$ and the recently proposed tensor
$V$~\cite{SoCQG2009} both fulfil the Lorentz covariant, future
pointing and non-spacelike requirements in a small sphere limit.
They are defined in empty space as follows:
\begin{eqnarray}
B_{\alpha\beta\xi\kappa}&:=&R_{\alpha\lambda\xi\sigma}R_{\beta}{}^{\lambda}{}_{\kappa}{}^{\sigma}
+R_{\alpha\lambda\kappa\sigma}R_{\beta}{}^{\lambda}{}_{\xi}{}^{\sigma}
-\frac{1}{8}g_{\alpha\beta}g_{\xi\kappa}\mathbf{R}^{2},\\
V_{\alpha\beta\xi\kappa}&:=&R_{\alpha\xi\lambda\sigma}R_{\beta\kappa}{}^{\lambda\sigma}
+R_{\alpha\kappa\lambda\sigma}R_{\beta\xi}{}^{\lambda\sigma}
+R_{\alpha\lambda\beta\sigma}R_{\xi}{}^{\lambda}{}_{\kappa}{}^{\sigma}
+R_{\alpha\lambda\beta\sigma}R_{\kappa}{}^{\lambda}{}_{\xi}{}^{\sigma}
-\frac{1}{8}g_{\alpha\beta}g_{\xi\kappa}\mathbf{R}^{2},\quad
\end{eqnarray}
where $\mathbf{R}^{2}=R_{\rho\tau\xi\kappa}R^{\rho\tau\xi\kappa}$,
Greek letters mean spacetime and the signature we use is $+2$. The
associated known energy-momentum density are
\begin{equation}
B_{\mu\lambda\sigma\tau}t^{\lambda}t^{\sigma}t^{\tau}
\equiv{}V_{\mu\lambda\sigma\tau}t^{\lambda}t^{\sigma}t^{\tau}
=(E_{ab}E^{ab}+H_{ab}H^{ab},2\epsilon_{cab}E^{a}{}_{d}H^{bd}),\label{29aSep2013}
\end{equation}
where Latin denotes spatial indices. The electric part $E_{ab}$
and magnetic part $H_{ab}$, are defined in terms of the Weyl
curvature~\cite{Carmeli}: $E_{ab}:=C_{ambn}t^{m}t^{n}$ and
$H_{ab}:=*C_{ambn}t^{m}t^{n}$, where $t^{m}$ is the timelike unit
vector and $*C_{\rho\tau\xi\kappa}$ indicates its dual for the
evaluation. Here we emphasize that both $B$ and $V$ are totally
traceless $\mathbf{t}_{\mu0\sigma}{}^{\sigma}=0$, which means
$\mathbf{t}_{\mu000}=\mathbf{t}_{\mu{}0ij}\delta^{ij}$, where
$\mathbf{t}$ can be replaced by $B$ or $V$. Moreover, the energy
component in (\ref{29aSep2013}) is non-negative definite for all
observers, which is well known, and the linear momentum component
is a kind of cross product between $E$ and $H$:
\begin{eqnarray}
\epsilon_{cab}E^{a}{}_{d}H^{bd}
&=&(\epsilon_{1ab}E^{a}{}_{d}H^{bd},\epsilon_{2ab}E^{a}{}_{d}H^{bd},\epsilon_{3ab}E^{a}{}_{d}H^{bd})\nonumber\\
&=&(E_{2a}H^{3a}-E_{3a}H^{2a},E_{3a}H^{1a}-E_{1a}H^{3a},E_{1a}H^{2a}-E_{2a}H^{1a})\nonumber\\
&=&(A_{x},A_{y},A_{z}),\label{30aDec2013}
\end{eqnarray}
where
$\vec{A}:=(E_{1a}\times{}H^{1a}+E_{2a}\times{}H^{2a}+E_{3a}\times{}H^{3a})$.
The cross product can be well-defined if we treat $E_{1a}$ as a
3-dimensional vector, explicitly $E_{1a}=(E_{11},E_{12},E_{13})$.
Similarly for $E_{2a}$, $E_{3a}$, $H_{1a}$, $H_{2a}$ and $H_{3a}$.
Referring to (\ref{30aDec2013}), the momentum magnitude can be
interpreted as follows
\begin{eqnarray}
|\epsilon_{cab}E^{a}{}_{d}H^{bd}|
&=&|E_{1a}\times{}H^{1a}+E_{2a}\times{}H^{2a}+E_{3a}\times{}H^{3a}|\nonumber\\
&\leq&|E_{1a}\times{}H^{1a}|+|E_{2a}\times{}H^{2a}|+|E_{3a}\times{}H^{3a}|\nonumber\\
&=&|E_{1a}||H_{1b}||\sin\theta_{1}|+|E_{2a}||H_{2b}||\sin\theta_{2}|+|E_{3a}||H_{3b}||\sin\theta_{3}|,
\end{eqnarray}
where $\theta_{1}$ is the angle between $E_{1a}$ and $H_{1a}$;
similarly for $\theta_{2}$ and $\theta_{3}$.

According to (\ref{29aSep2013}), both $B$ and $V$ have the same
Bel-Robinson type energy-momentum in a small sphere region, which
exhibits the desired causal relationship:
\begin{equation}
\mathbf{t}_{0000}-|\mathbf{t}_{000c}|=(E_{ab}E^{ab}+H_{ab}H^{ab})-|2\epsilon_{cab}E^{a}{}_{d}H^{bd}|\geq{}0,
\label{9aOct2013}
\end{equation}
and $\mathbf{t}$ can be either $B$ or $V$.  Here we consider two
more possibilities for the comparison with the energy and still
obtain the non-negative condition:
\begin{eqnarray}
(E_{ab}E^{ab}+H_{ab}H^{ab})+k_{1}(E_{ab}E^{ab}-H_{ab}H^{ab})\geq0,\quad{}&\Rightarrow&\quad{}
|k_{1}|\leq1,\label{22aOct2013}\\
(E_{ab}E^{ab}+H_{ab}H^{ab})+k_{2}E_{ab}H^{ab}\geq0,\quad{}&\Rightarrow&\quad{}
|k_{2}|\leq2.\label{22bOct2013}
\end{eqnarray}
The above two extra invariant terms come from
\begin{eqnarray}
R_{\alpha\beta\mu\nu}R^{\alpha\beta\mu\nu}=8(E_{ab}E^{ab}-H_{ab}H^{ab}),\quad{}
R_{\alpha\beta\mu\nu}*R^{\alpha\beta\mu\nu}=16E_{ab}H^{ab}.\label{30bDec2013}
\end{eqnarray}
These two terms are scalar and satisfy the Lorentz covariant
property. The first term can be classified as the energy density
(i.e., see (\ref{30cDec2013})$-$(\ref{30dDec2013})) and the second
as the momentum density (i.e., look
(\ref{30eDec2013})$-$(\ref{30fDec2013})). Moreover, the momentum
density $EH$ can be classified as a dot product between $E$ and
$H$:
\begin{eqnarray}
E_{ab}H^{ab}&=&E_{1a}H^{1a}+E_{2a}H^{2a}+E_{3a}H^{3a}\nonumber\\
&=&|E_{1a}||H_{1b}|\cos\theta_{1}+|E_{2a}||H_{2b}|\cos\theta_{2}+|E_{3a}||H_{3b}|\cos\theta_{3},
\end{eqnarray}
Combining the inequalities from (\ref{9aOct2013}) to
(\ref{22bOct2013})
\begin{equation}
(E_{ab}E^{ab}+H_{ab}H^{ab})+k_{1}(E_{ab}E^{ab}-H_{ab}H^{ab})
+k_{2}E_{ab}H^{ab}-|2\epsilon_{cab}E^{a}{}_{d}H^{bd}|
\geq{}0.\label{27aSep2013}
\end{equation}
Based on the argument from Szabados~\cite{Szabados}, the above
non-negative inequality should hold only if $k_{1}$ and $k_{2}$
are both zero. However, we can demonstrate that this is not true.
Let $|H_{Ia}|=\alpha_{I}|E_{Ia}|$ and $\alpha_{I}\geq0$, where
$I=1,2,3$, consider (\ref{27aSep2013}) again
\begin{eqnarray}
&&(E_{ab}E^{ab}+H_{ab}H^{ab})+k_{1}(E_{ab}E^{ab}-H_{ab}H^{ab})+k_{2}E_{ab}H^{ab}-2|\epsilon_{cab}E^{ad}H^{b}{}_{d}|\nonumber\\
&\geq&(E^{2}_{ab}+H^{2}_{ab})+k_{1}(E^{2}_{ab}-H^{2}_{ab})-|k_{2}||E_{ab}H^{ab}|
-2|\epsilon_{cab}E^{ad}H^{b}{}_{d}|\nonumber\\
&\geq&(1+k_{1})E^{2}_{1a}+(1-k_{1})H^{2}_{1a}
-|k_{2}||E_{1a}||H_{1b}||\cos\theta_{1}|-2|E_{1a}||H_{1b}||\sin\theta_{1}|
\nonumber\\
&&+(1+k_{1})E^{2}_{2a}+(1-k_{1})H^{2}_{2a}
-|k_{2}||E_{2a}||H_{2b}||\cos\theta_{2}|-2|E_{2a}||H_{2b}||\sin\theta_{2}|\nonumber\\
&&+(1+k_{1})E^{2}_{3a}+(1-k_{1})H^{2}_{3a}
-|k_{2}||E_{3a}||H_{3b}||\cos\theta_{3}|-2|E_{3a}||H_{3b}||\sin\theta_{3}|\nonumber\\
&=&\left\{ (1-\alpha_{I})^{2}\left[1
+\frac{k_{1}(1+\alpha_{I})}{(1-\alpha_{I})}\right]+2\alpha_{I}\left(1-\frac{1}{2}|k_{2}||\cos\theta_{I}|
-|\sin\theta_{I}|\right) \right\}E^{2}_{Ia}
\nonumber\\
&\geq&0, \label{3aDec2013}
\end{eqnarray}
provided that
\begin{eqnarray}
k_{1}\geq\frac{\alpha_{I}-1}{\alpha_{I}+1},\quad{}|k_{2}|\leq\frac{2(1-|\sin\theta_{I}|)}{|\cos\theta_{I}|}.
\end{eqnarray}
Thus (\ref{3aDec2013}) is non-negative for some non-vanishing
$k_{1}$ and $k_{2}$. The component with $k_{1}$ varies the energy
density, while the component with $k_{2}$ alters the momentum
value. One may question the purpose for this kind of modification,
but for the present discussion we note that we do not change the
energy-momentum relationship indicated in (\ref{9aOct2013})
through the introduction of the two terms multiplied by $k_{1}$
and $k_{2}$. The detailed physical consequences will be discussed
in section 3, i.e., see (\ref{5aOct2013}), (\ref{21aJan2013}) and
(\ref{14aMay2014})

Actually, we are repeating the same comparison with
Szabados~\cite{Szabados}. However, we have found a different
result; one that is strictly forbidden according to the conclusion
of Szabados's article. A natural question if (\ref{3aDec2013}) is
correct, is what are the allowed ranges for $k_{1}$ and $k_{2}$?
More precisely, looking at (\ref{27aSep2013}) again, we consider
what ranges for constants $k_{1}$ and $k_{2}$ may be selected such
that the Lorentz covariant and future directed non-spacelike
qualities can be kept.  For this purpose we use the 5 Petrov
types~\cite{Lobo} Riemann curvature for the verification. After
some simple algebra, we find a different results from
Szabados~\cite{Szabados}:
\begin{equation}
|k_{1}|\leq1, \quad{}|k_{2}|\leq2(1-|k_{1}|).\label{3bDec2013}
\end{equation}
This indicates that, in terms of a quasi-local energy-momentum
expression, $B$ and $V$ are not the only candidates that satisfy
the Lorentz covariant and future directed non-spacelike
requirements in a small sphere limit. There exists some relaxation
freedom for the modification, the detail will be discussed in
three cases in section 3. Here we list out the accompanied tensor
$S$ with $B$ or $V$ as follows:
\begin{equation}
S_{\alpha\beta\xi\kappa}=R_{\alpha\xi\lambda\sigma}R_{\beta\kappa}{}^{\lambda\sigma}
+R_{\alpha\kappa\lambda\sigma}R_{\beta\xi}{}^{\lambda\sigma}
+\frac{1}{4}g_{\alpha\beta}g_{\xi\kappa}\mathbf{R}^{2}.
\end{equation}

\section{Quasi-local energy-momentum}
We now examine the positive definite gravitational quasi-local
energy-momentum, which satisfies the Lorentz covariant and future
directed non-spacelike conditions.

Case (i): Consider a simple physical situation such that within a
small sphere limit we define: $\mathbf{t}+sS$, where $\mathbf{t}$
can be replaced by $B$ or $V$, and $s$ is a constant. For constant
time $t_{0}=0$, the energy-momentum in vacuum with radius $r$
\begin{equation}
2\kappa{\cal{}P}_{\mu}=\int_{t_{0}}(\mathbf{t}^{0}{}_{\mu\xi\kappa}+sS^{0}{}_{\mu\xi\kappa})x^{\xi}x^{\kappa}dV
=\frac{4\pi}{15}r^{5}(\mathbf{t}^{0}{}_{\mu{}ij}+sS^{0}{}_{\mu{}ij})\delta^{ij},\label{5aOct2013}
\end{equation}
where $\kappa=8\pi{}G/c^{4}$, $G$ is the Newtonian constant and
$c$ the speed of light. According to~\cite{Szabados}, the only
possibility is $s=0$ in order to produce the Lorentz covariant,
future pointing and non-spacelike properties. However, we can show
that there are some $s\neq0$ such that these properties are
preserved. As the 4-momentum of
$S_{0\mu{}ij}\delta^{ij}=-10(E^{2}_{ab}-H^{2}_{ab},0,0,0)$, we
only vary the energy and without affecting the momentum. After the
substitution, the energy for (\ref{5aOct2013}) is
\begin{equation}
-{\cal{P}}_{0}={\cal{}E}=\frac{2\pi}{15\kappa}r^{5}
\left[(E_{ab}E^{ab}+H_{ab}H^{ab})-10s(E_{ab}E^{ab}-H_{ab}H^{ab})\right],\label{19aOct2013}
\end{equation}
and the associated momentum is
${\cal{}P}_{c}=\frac{2\pi}{15\kappa}r^{5}(2\epsilon_{cab}E^{a}{}_{d}H^{bd})$.
Since the values of $E_{ab}$ and $H_{ab}$ can be arbitrary at a
given point, the sign of the energy component of $S$ is uncertain
and obviously $S$ affects the desired Bel-Robinson type
energy-momentum inequality: ${\cal{}E}\geq|\vec{{\cal{P}}}|$.
Previously, our preference was achieving a multiple of pure
Bel-Robinson type energy-momentum in a small
sphere~\cite{SoCQG2009}, and we thought the result in
(\ref{5aOct2013}) required $s=0$. However, we have now shown that
this is not true: we have found that certain linear combinations
of $\mathbf{t}$ and $S$ are legitimate. Comparing
(\ref{3aDec2013}) and (\ref{19aOct2013}), we observe that
$|k_{1}|=10|s|\leq1$ and $k_{2}=0$ produce results that satisfy
the non-negative energy, Lorentz covariant and future directed
non-spacelike requirements. Here we give a remark: previously we
thought both Einstein $\mathbf{t}^{E}_{\alpha\beta}$ and
Landau-Lifshitz $\mathbf{t}^{LL}_{\alpha\beta}$ pseudotensors
cannot give the positive (i.e., causal) definite quasi-local
energy in Riemann normal coordinates~\cite{1stSoCQG2009}:
\begin{equation}
{\mathbf{t}}^{E}_{\alpha\beta}
=\frac{2}{9}\left(B_{\alpha\beta\xi\kappa}-\frac{1}{4}S_{\alpha\beta\xi\kappa}\right)x^{\xi}x^{\kappa},\quad{}
{\mathbf{t}}^{LL}_{\alpha\beta}
=\frac{7}{18}\left(B_{\alpha\beta\xi\kappa}+\frac{1}{14}S_{\alpha\beta\xi\kappa}\right)x^{\xi}x^{\kappa}.
\label{15aMay2014}
\end{equation}
This implies that the Landau-Lifshitz pseudotensor (i.e.,
corresponding $|s|=\frac{1}{14}<\frac{1}{10}$) is a suitable
candidate for the Lorentz covariant and future directed
non-spacelike requirements, while Einstein pseudotensor does not
(i.e., associated $|s|=\frac{1}{4}>\frac{1}{10}$).

Case (ii): Evaluate the energy-momentum in a small ellipsoid,
replacing $\mathbf{t}$ by $B$ or $V$. Consider a simple dimension
$(a,b,c)=(\sqrt{1+\Delta},1,1)r_{0}$ for non-zero $|\Delta|<<1$
and $r_{0}$ finite.  For constant time $t_{0}=0$, the
corresponding 4-momentum are
\begin{equation}
2\kappa{\cal{}P}_{\mu}=\int_{t_{0}}\mathbf{t}^{0}{}_{\mu{}ij}x^{i}x^{j}dV
=\frac{4\pi}{15}(\mathbf{t}^{0}{}_{\mu{}ij}\delta^{ij}+\Delta\mathbf{t}^{0}{}_{\mu{}11})r^{5}_{0}\sqrt{1+\Delta}.
\label{21aJan2013}
\end{equation}
Here we list out the energy component for $B$ and $V$
\begin{eqnarray}
B_{0011}&=&E_{ab}E^{ab}+H_{ab}H^{ab}-2E_{1a}E^{1a}-2H_{1a}H^{1a},\label{30cDec2013}\\
V_{0011}&=&3E_{ab}E^{ab}-H_{ab}H^{ab}-8E_{1a}E^{1a}+4H_{1a}H^{1a},\label{30dDec2013}
\end{eqnarray}
and the associated momenta are
\begin{eqnarray}
&&B_{0c11}=2\epsilon_{cab}(E^{ad}H^{b}{}_{d}-2E^{a}{}_{1}H^{b}{}_{1}),\label{30eDec2013}\\
&&V_{0c11}=2\epsilon_{1ab}(E^{ad}H^{b}{}_{d}-2E^{a}{}_{1}H^{b}{}_{1},
2E^{a}{}_{1}H^{b}{}_{2}-4E^{a}{}_{2}H^{b}{}_{1},
2E^{a}{}_{1}H^{b}{}_{3}-4E^{a}{}_{3}H^{b}{}_{1}).\label{30fDec2013}\quad\quad
\end{eqnarray}
Looking at (\ref{21aJan2013}), $\Delta\mathbf{t}^{0}{}_{\mu{}11}$
varies the energy and momentum of
$\mathbf{t}^{0}{}_{\mu{}ij}\delta^{ij}$ simultaneously, i.e.,
making it analogous with (\ref{3aDec2013}):
$k_{1}\neq{}0\neq{}k_{2}$. Using the 5 Petrov types Riemann
curvature to compare the energy and momentum in
(\ref{21aJan2013}), we find that if $\mathbf{t}$ is replaced by
$B$ the Lorentz covariant and future directed non-spacelike
properties require $\Delta\in(-1,1]$. Similarly, if we replace
$\mathbf{t}$ by $V$, it is also true provided
$\Delta\in[-\frac{1}{3},\frac{1}{5}]$. However, as far as the
quasi-local small 2-surface is concerned, practically, we only
need the non-zero $\Delta$ to be sufficiently small. Therefore,
the result in (\ref{21aJan2013}), a linear combination for
$\mathbf{t}^{0}{}_{\mu{}ij}\delta^{ij}$ with an extra
$\mathbf{t}^{0}{}_{\mu{}11}$, is a physically reasonable candidate
for describing the quasi-local energy-momentum.

Case (iii): Demonstrate the total energy-momentum on a gravitating
system by an external universe, i.e., transferring the
gravitational field energy from Jupiter to Io. Referring to second
equation of (\ref{15aMay2014}), evaluate the energy-momentum for
Landau-Lifshitz pseudotensor in a small ellipsoid. It is natural
to consider a 2-surface ellipsoid instead of a 2-surface sphere
because Jupiter deformed Io from being a perfect sphere through
the tidal force. In reality, it is slightly deformed and it suits
the quasi-local small 2-surface limit. The detail is follows.
Again let $(a,b,c)=(\sqrt{1+\Delta},1,1)a_{0}$, constant time
$t_{0}=0$ and the 4-momentum are
\begin{equation}
2\kappa{\cal{}P}^{LL}_{\mu}
=\frac{14\pi}{135}\left[(B^{0}{}_{\mu{}ij}+sS^{0}{}_{\mu{}ij})\delta^{ij}
+\Delta(B^{0}{}_{\mu{}11}+sS^{0}{}_{\mu{}11})\right]
a^{5}_{0}\sqrt{1+\Delta},\label{14aMay2014}
\end{equation}
where $s=\frac{1}{14}$, energy from
$S_{0011}=-2(E^{2}_{ab}+2E^{2}_{1a}-H^{2}_{ab}-2H^{2}_{1a})$ and
momentum from
$S_{0c11}=4(0,E_{1a}H_{3}{}^{a}+E_{3a}H_{1}{}^{a},-E_{1a}H_{2}{}^{a}-E_{2a}H_{1}{}^{a})$.
Looking at (\ref{14aMay2014}) for the 4-momentum, we observed that
the interval for $\Delta\in[-\frac{1}{3},\frac{1}{5}]$ satisfies
the requirements for the Lorentz covariant and future directed
non-spacelike. Recall $\frac{GM}{c^{2}r}=3.4\times10^{-9}$ which
is small compare to unity (i.e., weak gravity limit), where
$M=1.90\times10^{27}$kg denotes the mass of Jupiter,
$r=4.2\times10^{5}$km means the separation between Jupiter and Io.
The physical dimension for Io is $(x,y,z)=(3660.0, 3637.4,3630.6)$
in kilometer. Using our notation: $a=\sqrt{1+\Delta}\,a_{0}$,
$b\simeq{}c\simeq{}a_{0}$, where $a_{0}=1817$km and
$\Delta=0.0144$. Indeed this ellipsoid is a little bit deformed
from a perfect sphere. In our case, the volume element of Io is
the quasi-local 2-surface for evaluating the energy-momentum
values.  Note that the density of Io is
$M_{Io}=8.93\times10^{22}$kg.  Let's use the Schwarzchild metric
in spherical coordinates~(see \S31.2 in \cite{MTW}) for a simple
test. Certainly, there is no momentum since we are dealing with a
static spacetime. The non-vanishing Riemann curvatures are
$R_{\hat{t}\hat{r}\hat{t}\hat{r}}=-R_{\hat{\theta}\hat{\phi}\hat{\theta}\hat{\phi}}
=-\frac{2GM}{c^{2}r^{3}}$ and
$R_{\hat{t}\hat{\theta}\hat{t}\hat{\theta}}=R_{\hat{t}\hat{\phi}\hat{t}\hat{\phi}}
=-R_{\hat{r}\hat{\theta}\hat{r}\hat{\theta}}=-R_{\hat{r}\hat{\phi}\hat{r}\hat{\phi}}=\frac{GM}{c^{2}r^{3}}$.
Substitute into (\ref{14aMay2014}) and thence the total
energy-momentum complex (see (29) in \cite{SoNesterCQG2009} and
(45) in \cite{Purdue}) is
\begin{eqnarray}
{\cal{T}}^{LL}_{00}&=&T^{LL}_{00}+(2\kappa)^{-1}\mathbf{t}^{LL}_{00}\nonumber\\
&=&M_{Io}
+\frac{14\pi{}G^{2}M^{2}}{45\kappa{}c^{4}r^{6}}\left[(1-10s)+\frac{\Delta}{3}(1-10s)\right]a^{5}_{0}\sqrt{1+\Delta}\nonumber\\
&=&1.11M_{Io}.
\end{eqnarray}
Note that the extra amount of energy received from Jupiter is
small but significant.

\section{Conclusion}
To describe the positive quasi-local energy-momentum expression,
the Bel-Robinson tensor $B$ and tensor $V$ are suitable because
both of them give the Bel-Robinson type energy-momentum in a small
sphere region. In the past, it has seemed that only this
Bel-Robinson type energy-momentum can manage this specific task:
Lorentz covariant, future pointing and non-spacelike. That
particular restriction cannot allow even a small amount of energy
to be subtracted from this Bel-Robinson type energy-momentum.
After some careful comparison and using the 5 Petrov type Riemann
curvature for the verification, we have discovered that the
Bel-Robinson type energy-momentum implies Lorentz covariant and
future directed non-spacelike properties; but the converse is not
true. We find that there exists a certain relaxation freedom such
that one can (i): add an extra tensor $S$ with $B$ or $V$ in a
quasi-local small sphere limit, or (ii): directly evaluate $B$ or
$V$ in a small ellipsoid region, (iii): Using the Landau-Lifshitz
pseudotensor to calculate the total energy, refer to the
Schwarzchild metric, in a small ellipsoid region.

Previously, we thought there are only two classical
energy-momentum expressions, Papapetrou pseudotensor and
tetrad-teleparallel energy-momentum gauge current expression, that
contribute the desired Lorentz covariant and future directed
non-spacelike requirements. Now, we have to add one more:
Landau-Lifshitz pseudotensor in Riemann normal coordinates.

\section*{Acknowledgment}
The author would like to thank Dr. Peter Dobson, Professor
Emeritus, HKUST, for reading the manuscript and providing some
helpful comments. This work was supported by NSC 99-2811-M-008-021
and NSC 100-2811-M-008-063.

%\bibitem{Deser}
%Deser S, Franklin J S and Seminaea D 1999 {\it Class. Quantum
%Grav.} {\bf 16} 2815

\end{document}